\documentclass[manuscript,screen,nonacm]{acmart}

\AtBeginDocument{%
  \providecommand\BibTeX{{%
    \normalfont B\kern-0.5em{\scshape i\kern-0.25em b}\kern-0.8em\TeX}}}

\begin{document}

\title{CULLING MISINFORMATION FROM GEN AI: Toward Ethical Curation and Refinement}


\author{Prerana Khatiwada}
\affiliation{%
  \institution{University of Delaware}
  \city{Newark}
  \country{USA}
  }
\email{preranak@udel.edu
}
\author{Grace Donaher}
\affiliation{%
   \institution{University of Delaware}
  \city{Newark}
  \country{USA}
  }
\email{gdonaher@udel.edu}

\author{Jasymyn Navarro}
\affiliation{%
 \institution{University of Delaware}
  \city{Newark}
  \country{USA}
  }
\email{jasnav@udel.edu}

\author{Lokesh Bhatta}
\affiliation{%
 \institution{Wilmington University}
  \city{New Castle}
  \country{USA}
  }
\email{lbhatta001@my.wilmu.edu}

\renewcommand{\shortauthors}{Khatiwada, et al.}








\begin{abstract}
While Artificial Intelligence (AI) is not a new field, recent developments, especially with the release of generative tools like ChatGPT, have brought it to the forefront of the minds of industry workers and academic folk alike. There is currently much talk about AI and its ability to reshape many everyday processes as we know them through automation. It also allows users to expand their ideas by suggesting things they may not have thought of on their own and provides easier access to information. However, not all of the changes this technology will bring or has brought so far are positive; this is why it’s extremely important for all modern people to recognize and understand the risks before using these tools and allowing them to cause harm. This work takes a position on better understanding many equity concerns and the spread of misinformation that result from new AI, in this case, specifically ChatGPT and deepfakes, and encouraging collaboration with law enforcement, developers, and users to reduce harm. Considering many academic sources, it warns against these issues, analyzing their cause and impact in fields including healthcare, education, science, academia, retail, and finance. Lastly, we propose a set of future-facing guidelines and policy considerations to solve these issues while still enabling innovation in these fields, this responsibility falling upon users, developers, and government entities.

\end{abstract}





\maketitle

\section{Introduction}
While Artificial Intelligence (AI) has been studied since the 1960s, Generative AI—capable of creating human-like text, images, and other content—has only recently transitioned from research labs to widespread public use. AI has brought profound change to many industries, transforming fields such as healthcare \cite{ 2saleem2023ethics}, education \cite{1bond2024meta}, retail \cite{3lee2024human, 6goti2023artificial}, finance \cite{8kalyani2023artificial}, and more. AI refers to a machine's ability to perform tasks that typically require human intelligence, such as learning, decision-making, problem-solving, and/or perception \cite{8kalyani2023artificial}. AI can be divided into three categories: computer vision (CV), natural language processing (NLP), and other Machine Learning (ML) systems \cite{6goti2023artificial}. Computer vision tasks include object detection, image classification, and medical image processing. While the task NLPs are often used for is text classification, as many of these models are based on pre-trained models \cite{7zheng2023ai}. Machine learning, however, utilizes statistical algorithms to interpret patterns from data and make predictions based on those patterns \cite{8kalyani2023artificial}.
Despite the apparent opportunities that come with this technology, the uptake of its usage has received a response from public discourse \cite{1bond2024meta}. With the appearance of software such as ChatGPT, citizens are concerned about readiness, ethics, privacy, impact, and the need for regulation, research, and training required to keep up with the rapid speed of AI's development. Especially now, the release of ChatGPT's models 3 and 3.5 enabled the transformer to create coherent and textually relevant responses to the prompts given to it \cite{4elkhatat2023evaluating}. 

While appearing promising, due to a lack of regulation, this technology is still vulnerable to abuse and misuse \cite{ 11godde2023swot}. Some of the misuse of this technology is due to bad actors taking advantage of the technology to spread incorrect or inappropriate content of individuals through deepfakes \cite{13juefei2022countering}. For example, AI systems can only be as objective as the data used to train them. If that data is biased, then the system is biased \cite{ 2saleem2023ethics}. So, when implemented in areas such as healthcare, this can lead to disparities in different patient populations, raising safety and equity concerns. Thus far, deepfake technology has been used to target many prominent female figures, such as actresses or celebrities \cite{13juefei2022countering}, for example, the ability to swap faces. Additionally, this could lead to future issues outside our line of sight.
Another example is that international governments have already created regulations like the European EU AI Act. Additionally, Australia has established a task force to develop a framework for AI usage in schools \cite{1bond2024meta}. Many other experts continue to call for some form of regulation, which this paper aims to provide aid in addressing.

Two primary examples of modern generative AI technology that this paper highlights are ChatGPT and deep fakes, which have respective abilities to amplify the spread of false information, whether due to an error within the program or intentional misuse. 
This paper takes the position that while generative AI offers significant opportunities—such as improving efficiency and expanding access to information—it also introduces critical risks that must be addressed. We present arguments highlighting both the potential and the pitfalls of technologies like ChatGPT and deepfakes, focusing especially on their role in amplifying misinformation and reinforcing social inequities. Through this lens, we aim to inform and guide future research, policy, and collaborative interventions.

The primary contributions of this paper are twofold: (1) it analyzes the opportunities presented by AI, such as enhancing the efficiency of routine tasks; and (2) it highlights the emerging risks associated with its widespread integration, including the spread of misinformation and the reinforcement of social inequities. Misinformation being spread, one primary concern of many technical experts, can appear much more credible because of recent advancements in this field. Other concerns revolve around its impacts on specific groups, such as racial minorities or females, and how this technology is capable of amplifying equity concerns. Lastly, we aim to provide a solution that can effectively mitigate misinformation and equity concerns, calling upon users, developers, and law enforcement to do their part in stopping the spread of false information and harm to individuals that AI biases and bad actors can inflict.

\section{Related Works}
To contextualize this paper, it’s essential to recognize the current role of AI and how it has shaped user-generated content (UGC) to date. AI is not only impacting the technology field but is also seeing integration in many different fields
as well as providing assistance to individuals for personal use.

\subsection{MODERN PRESENCE OF AI }
While AI has been an area of interest for many people for decades, recent advances in AI have made it applicable to many industries, including healthcare \cite{2saleem2023ethics}. Common themes of its presence are that it revolutionizes processes and enhances efficiency \cite{ 2saleem2023ethics}, one of its benefits being its ability to quickly process large amounts of data. AI was able to advance diagnostic and treatment planning in the medical field \cite{2saleem2023ethics}. Machine learning (ML) algorithms can cross-analyze large-scale datasets, including information like medical history, lab results, and images, to find trends in existing data to forecast potential health problems \cite{2saleem2023ethics}. This allows them to diagnose much faster and more accurately. AI also has many other implications for healthcare, including automating systems pertaining to patient management and healthcare delivery. Using AI systems in patient management can help with early intervention, reduce adverse effects, and provide more personalized care. For example, these systems can monitor patient vitals in real-time, alerting physicians of issues before they become egregious. Additionally, healthcare delivery can become more efficient with the automation of appointment scheduling and prescription refills. While there are many great examples of modern applications of AI, the benefits only begin with healthcare. In other fields, such as the retail industry, there is an incredible increase in integration with AI technology. This can be seen in the promising applications in fashion design. The AI-assisted design process can use AI-powered search engines to predict future fashion trends and allow designers to develop new designs inspired by these trends \cite{3lee2024human}. Additionally, AI can manage clients and items, making it easy for businesses to understand every aspect of each product's storage, distribution, and sale. By using social media and customer reviews to monitor how consumers feel about a product's different aspects, the company can modify future products to meet consumer needs and desires \cite{6goti2023artificial}.

\subsection{AI’S IMPACT ON USER-GENERATED CONTENT}
AI has opened numerous possibilities in academia, including creating a personalized learning experience, AI-based tutoring systems, detecting plagiarism, and giving personalized feedback to students \cite{1bond2024meta}. For example, the AI-powered tool ChatGPT, which allows students to use prompts to generate customized content, has emerged \cite{1bond2024meta}. While ChatGPT can serve students by allowing them easy access to answers to specific questions they may have regarding the content they are learning or assistance in other academic tasks, it has also raised concerns about the originality, appropriateness, and accuracy of its use in an academic setting \cite{4elkhatat2023evaluating}. The issue with originality comes from a concern about plagiarism. ChatGPT employs a pre-trained dataset to create responses to prompts. Many experts claim that ChatGPT does not follow the ICMJE guidelines for authorship. After conducting plagiarism checks on text generated by ChatGPT, they found many examples of direct, paraphrasing, and source-based plagiarism \cite{2saleem2023ethics}. Additionally, the concern about the appropriateness of ChatGPT use by a student comes from how it impedes a teacher/professor from evaluating what the student has learned and is competent at explaining. However, while the technology is much more successful than its predecessors, according to OpenAI, its knowledge is limited to data obtained before September 2021. This can cause inaccuracy issues for anyone using ChatGPT on a subject requiring information post the cutoff date. It is evident that while ChatGPT has the ability to revolutionize academia, it is capable of undermining academic integrity. Not only has AI proved to be helpful in academic or industrial settings, but it is also changing the way art is created. Recent advances in text-to-image generation now allow Large Language Models (LLMs) to retrieve pre-existing images and even create novel images \cite{5pavez2023advanced}.

\section{Ethical Framework and Position Statement}
With the rise of Artificial Intelligence-Generated Content (AIGC) technology, there is a rush to find its implementations in industries \cite{2saleem2023ethics, 3lee2024human, 6goti2023artificial, 8kalyani2023artificial} and academia \cite{1bond2024meta} alike. This tech brings new opportunities for automation. However, it appears that these fields are presented with new opportunities and challenges these new technologies bring in \cite{8kalyani2023artificial}. While many modern artificial intelligence models exhibit incredible language-processing capabilities, such as ChatGPT’s ability to generate contextually relevant and human-like responses, it’s important to note that they lack proper contextual understanding and comprehension \cite{10florindo2023chatgpt}. 

\subsection{Ethical Concerns of ChatGPT }
The aforementioned ChatGPT is a large language model developed by OpenAI, an AI  research and development company based in San Francisco, California, that was released at the end of 2022 \cite{10florindo2023chatgpt}. This model is based on the Generative Pre-trained Transformer architecture, represented by the GPT in its name. The modern influence of ChatGPT ranges from providing quick access to information, assisting in writing medical and scientific papers, and even interpreting complex datasets \cite{ 11godde2023swot}. ChatGPT offers a lot of opportunities to authors in many fields; for example, it assists non-native English speakers, helping to improve their English proficiency or helping with data analysis; with its ability to process large volumes of data, it can expedite these processes, saving valuable time and resources \cite{10florindo2023chatgpt}. However, there are still risks when using this technology, a prominent risk being inaccuracy. Since the chatbot's text appears credible, this amplifies the risk of spreading false information, leading to public confusion \cite{10florindo2023chatgpt} and potentially compromising safety, mainly when used in the medical field \cite{11godde2023swot}. 

An example of misinformation spread by ChatGPT is biased. AI chatbots learn from large datasets, and if that data is biased in any way, the system has bias \cite{2saleem2023ethics}. Not only does ChatGPT pose a threat of inaccuracy for the user, but it also raises concerns about plagiarism. Any content developed by AI is based on the pre-existing data it has access to, making it plagiarism or now being dubbed “AI-giarism” \cite{14test}. This has spurred a debate about the transparency of AI usage in writing. While some have suggested citing AI tools as authors may be in order, and certain scientists have done so in their publications, these tools cannot take credit for the contributions made \cite{10florindo2023chatgpt}. While many teachers in higher education believe they can use plagiarism detection tools to catch students who use ChatGPT or other Chatbots in their writing assignments, it is unfortunately not so easy to detect. AI tools do not always steal text directly from different sources; instead, they create their own based on the expansive resource of ideas on the internet \cite{ 14test}. And since AI will only improve with time, it will become increasingly more challenging to detect. 

\subsection{Ethical Concerns of Deepfake Technology}
A variant of machine learning applications that have hit mainstream media is deep fakes and their obscurification of online trust. Deep fakes can be recognized as using machine learning to alter an image or video, using a person’s visage to make them say or do something they may not have done \cite{12de2021distinct}. This form of media is not inherently malicious, but it’s probable to compare it to a loaded gun in the wrong hands as it can be used for severe harm \cite{13juefei2022countering}. Consider first the merit of integrating deep fake tech as an agent of innovation in fields such as medicine. Patients with ALS suffer from losing part of their identity when they can no longer use their voices. The effect is devastating since it seems like one has lost a part of oneself. Regenerative capabilities of deepfake processing return this part of the victims, allowing those who lost their voice to hear them again \cite{12de2021distinct}. Communicating with text-to-speech is available but impersonal compared to hearing a loved one’s voice.

It’s evident that this technology is pliable for many uses, making it powerful. Instances of abuse using such power emphasize the ethical concerns associated with it. Particularly in media, deep fakes take the form of humorous parodies, played on a tone of satirical commentary, but at the deep end, are created with intent for harm. Early malicious uses of deepfakes included faces of female celebrities participating in sexual activities by pasting their faces on adult actors \cite{13juefei2022countering}. The defiling of celebrity images became such a trend that it gained much traction on Reddit in 2017, as a user posted various videos featuring multiple female actresses \cite{12de2021distinct}. What follows are forged videos of public figures participating in depravity, which is a blow to their image and reputation. Using deepfake technology to manipulate a person’s body or face to do something they did not is ethically disruptive, harming their sense of self and social identity \cite{12de2021distinct}. 

The scope of those affected is not limited to celebrities, but any public figure or person on the web. Both the victims of imitation and the users who consume or spread rumors fall for the deception of bad actors. The broader harm for online users is the psychological influence they come under when they take false information as truth. Political candidates who fall victim to deepfakes share in the harm that comes with them being presented in a potentially unsavory way unwillingly \cite{ 12de2021distinct}. The harm to victims’ sense of self, the spread of misinformation, and the ability to manipulate others’ reputations by bad actors all contribute to the necessity of policy surrounding such technology.

\section{Proposed Argument}
A future shaped by Generative AI calls for a bold rethinking of how we identify, govern, and educate against misinformation—approaches that go far beyond platform moderation or reactive legislation. As generative models become more sophisticated, policy must shift from simply chasing bad actors to anticipating misuse at the infrastructure and interactional levels. How we approach policy surrounding this topic will involve collaboration from policymakers at the online administrative, municipal, and federal levels. A truly multi-faceted approach must aim to curb the spread of misinformation by AI, put preventative measures in place to reduce future biases, and define what legislation should be passed to mitigate bad actors from using this technology wrongfully.

First, we propose the creation of \textbf{``Civic Model Registries''}—a mandatory, public-facing system where all large-scale generative models must be logged with metadata detailing training data provenance, known biases, update histories, and transparency scores. Much like nutritional labels, these registries would inform journalists, researchers, and users about a model's risk profile before its outputs circulate widely. No such centralized civic infrastructure currently exists, yet it is critical to transparency and accountability.

Preventative measures in places online where generated AI spreads misinformation to users are necessary to impede the flow before the situation becomes more dire.
Without timely intervention, the scale and credibility of such content may reach a point of irreversible public harm. One branch of this would include increased moderation on platforms. Similar to existing moderation, posted content that does not fit standards would be removed; however, future moderation would scan for misinformation by generating content using AI, not just bad decorum. For example, moderation systems might themselves generate alternate versions of posts using AI to evaluate intent, deception, or semantic manipulation. Importantly, AI-based moderation already enjoys growing support from users—particularly among liberal-majority demographics on platforms like Twitter\cite{15wang2023factors}.
So, integrating systems to weed out these faults would be theoretically accepted easily. Admins of these websites could also play a part in preventing Gen AI from being misinterpreted altogether. One of the most common ways misinformation spreads online is from users mistaking Deepfakes for genuine photos or videos \cite{12de2021distinct}. Preventing the spread starts at the source, meaning identifying what content an AI generates before it’s posted. This process may look like enforcing creatives to embed ‘watermarks’
or machine-detectable tags in all AI-generated content before uploading it, and then, when detected, the website can label the post with contextual information.
\looseness -1

Thus, second, we propose advancing content-level moderation into \textbf{“semantic provenance tracking”}—a next-generation system in which AI-generated outputs embed a traceable semantic fingerprint akin to a cryptographic hash of the model, prompt, and output context. This approach enables not only detection of AI-generated content but also lineage tracking of ideas, phrases, and framing patterns across languages, communities, and campaigns. Unlike current watermarking efforts, semantic provenance allows researchers and platform moderators to identify coordinated manipulation strategies over time and space—shifting from single-instance detection to systemic tracing of influence.

Third, we advocate for the development of \textbf{``counter-generative systems''}: AI models explicitly trained to detect, decode, and counteract persuasion tactics in real time. These models would not merely flag misinformation but proactively generate “cognitive countermeasures”—alternative framings, explanatory content, or neutral summaries that help users recognize manipulation without relying on content takedown. This shifts the paradigm from suppression to empowerment, giving users the tools to understand when and how they are being influenced, while preserving free expression.
\looseness -1

Policy frameworks restricting various uses for gen AI content at the municipal and federal levels are equally crucial to ensure accountability and disincentivize malicious use. Legislation is integral for procedures to follow in instances of abuse online using this machine learning technology. Laws create an expectation for both enforcers and citizens on what is acceptable without causing harm. The progression of laws addressing misuse has spurred around the U.S., as several states ratify their AI laws. Other states should follow suit, referring to laws that deter deepfakes, such as the Nonconsensual Pornography law in Virginia and the criminalization of deepfake video publication with intent of harm in Texas and California \cite{ 13juefei2022countering}. Such laws prevent harm to public figures, such as political candidates, who are at high risk of deteriorating their public image if tampered with by those with opposing views. However, there remains a critical gap in national and international policy coherence. Broader adoption of these legislative models could protect high-risk individuals such as political candidates, activists, and public figures whose reputations and credibility are particularly vulnerable to generative manipulation.


Finally, we propose embedding \textbf{``AI literacy modules''} directly into public-facing interfaces—social media platforms, search engines, and news aggregators—rather than relegating education to external media literacy campaigns. Platforms could include brief, transparent indicators such as: “This summary was generated by GPT-5, trained primarily on Western news sources from 2021–2023.” This low-friction disclosure would anchor user expectations and foster reflective consumption without demanding technical expertise.

Ultimately, these proposals represent a shift from reactive moderation to proactive infrastructure—an ecosystem where misinformation is curbed through traceability, accountability, and user empowerment. What bad actors do with generative media must be clearly defined in the eyes of the law, but that alone is not enough. Corralling AI-generated content only works if it can be reliably identified, tracked, and contextualized. Both law enforcement and everyday users must be made aware that the content they encounter may be false, manipulative, or synthetic in origin. The psychological effects of engaging with persuasive misinformation are real. Unless we equip the public with the ability to decode it, the erosion of shared truth will continue to deepen. Generative AI doesn’t merely call for stronger content filters; it demands an urgent rethinking of the systems of trust through which information circulates.

\section{FUTURE WORK AND LIMITATIONS}
The preceding research covered implied ethical considerations surrounding two types of generative AI technology: chat-based AI (such as ChatGPT) and deepfake media. It divulged the harmful effects that bad actors could use them for, or even the innate ethical wrongness of using them, because they spread misinformation. In turn, a proposal for action to address the lack of policy in the AI landscape was outlined. This looked like increased moderation on platforms where misinformation caused by AI often takes precedence, creating rumors and purging when necessary. In addition, identifies which content AI created at its source and brings back context to users. At a larger scope, we identified the need for action by government agencies to corral the leniencies in what was legally allowed but ethically wrong. Strict definitions of legally generated content are necessary in a landscape of limitless creativity.

Findings from this work are limited in scope, and they cannot divulge what policies or actions were effective so many years down the line. Generative AI content may not be new, but its popularity only recently soared. This limits the number of case studies for us to analyze and, in turn, prevents our complete understanding of what can be done. It should also be noted that the ethical considerations covered are supported by controversial evidence, meaning it’s hard to determine if the tech is inherently good or bad. The lack of a stance in our opinions is restricted due to the ongoing debate by those who benefit and are harmed by AI. 
Going forward, researchers should work towards refining generative AI systems to make them more impartial to biases. Biases existing in the data on which AI is trained are a large contributor to their perpetuation of inequity. This manifests as a strengthening of stereotypes and a lack of influence from minorities. The difficulties in executing this may show when trying to balance how much influence each characteristic of the data they include has. We should examine how these effects perpetuate the spread of misinformation and other contributing factors that may not be as intentional. Refining the training of generative AI models is a proactive approach to preventing harmful side effects in the public domain and builds a less ambivalent machine.

\section{CONCLUSION}
Applications of Generative AI content are undeniably a powerful prospect for how industries will function. The essential ethical dilemma that surrounds this technology is the power imbalance of innovation and harm done intentionally or not. A lack of preventative measures nerfing unsavory actions on the web bleeds misinformation into public spaces and can hurt the individuals targeted. This can mean victims in a distasteful deepfake or users’ perspectives after taking information made by a biased chatbot as fact. However, if all parties involved maintain the precedence of policy around gen AI and filter out its existence, then the utility of these applications can expedite processes. Sectors such as healthcare, education, and retail may benefit from leaning on the generative powers of machine learning in areas such as data processing, automation, and idea generation. Recent advancements, including models like ChatGPT 3-4, have highlighted the transformative potential of AI. While it brings positive changes, there are concerns about equity, misinformation, and potential harm to online users. Recognizing these risks, this paper emphasizes the importance of understanding and addressing issues related to generative AI, specifically ChatGPT and deepfakes. The work also provides future guidelines to balance innovation and accountability, emphasizing the shared responsibility of users, developers/admins, and government entities in creating a positive AI landscape.

\section{Acknowledgments}
We want to acknowledge that this work originated as a class assignment for the course Computer Ethics and Society. We extend our heartfelt gratitude to our advisor, Matthew Louis Mauriello, for his constructive feedback and guidance throughout the writing process. We also sincerely thank our class students for their valuable insights, collaborative spirit, and dedication, which significantly shaped this early work.

\newpage 

\bibliographystyle{ACM-Reference-Format}
\bibliography{sample-base}

\end{document}